# Black holes were born

Jeroen van Dongen, University of Amsterdam

*Abstract:*

*In 1939 Albert Einstein wrote a technical article that argued against the possibility that a star can be contracted to a single point: particles making up the star would end up rotating at velocities that were too high. In the same year, Robert Oppenheimer, together with his student Hartland Snyder, drew an apparently exactly opposite conclusion: that when a sufficiently heavy star runs out of nuclear fuel, it will collapse into an infinitely dense point, closed off from the rest of the universe. Both Oppenheimer and Einstein would soon be preoccupied by choices of an altogether different nature; and, again, set out on a different course.*

Black holes were born in 1939. In that year, J. Robert Oppenheimer and his student Hartland Snyder published their article *On continued gravitational contraction*. It began by observing that after a star burns through its nuclear fusion energies, the pressure of its radiation will no longer be able to counter the pull of its gravitational field. Oppenheimer and Snyder found that for a heavy star only one option remains: it can only contract. And that contraction will not halt until the star is reduced to a mere infinitesimal point. The theory of relativity, they argued, ensures that, eventually, the star "tends to close itself off from any communication with a distant observer; only its gravitational field persists."[1]

      Ideas that anticipated Oppenheimer and Snyder's insight were already circulating before 1939. The most important equation they used had been written down in 1916 by Karl Schwarzschild, a German astronomer who had been closely following the development of Albert Einstein's theory of relativity. Schwarzschild had determined how the gravitational field around a star should look in the new theory. To make his calculations easier, he had modeled the star as an infinitely small point. He hit upon a strange circumstance: at a small distance just outside this central mass, the gravitational field appeared to be infinitely large. Yet, for a mass that was comparable to the sun's, this distance was nearly zero. Furthermore, an infinity in the field could be expected as a mathematical artefact of such a naïve point source model, and Schwarzschild seemed not particularly concerned about it (Schwarzschild 1916a, 1916b). In any case, the small distance is now usually referred to as a star's 'Schwarzschild radius'.

      Many authors studied Schwarzschild's stellar model.[2] It offered the possibility to describe precisely the trajectories of planets and comets, or, for example, to determine how much the light emitted by a faraway object would be deflected when it grazes the surface of the sun. That calculation was instrumental for the British eclipse expeditions of 1919, which confirmed the

---

[1] J. Robert Oppenheimer and Hartland Snyder, 'On Continued Gravitational Contraction', *Physical Review*, 56 (1939), 455-459, on p. 456.
[2] For discussion of the early literature on the Schwarzschild spacetime, see Jean Eisenstaedt, 'Histoire et singularités de la solution de Schwarzschild (1915-1923)', *Archive for History of Exact Sciences*, 27 (1982), pp. 157-198; Jean Eisenstaedt, 'Trajectoires et impasses de la solution de Schwarzschild', *Archive for History of Exact Sciences*, 37 (1987), 275-357.

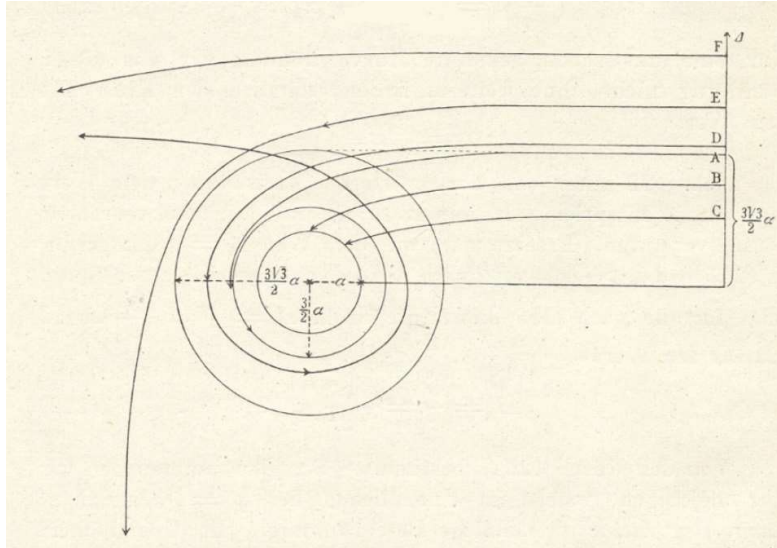

Figure 1. Trajectories of light at different impact parameters in a Schwarzschild spacetime before the black hole interpretation was introduced. As this illustration documents, anything that falls inwards was expected never to pass the 'Schwarzschild radius', here indicated by the distance $\alpha$. Taken from: Max von Laue, *Die Relativitätstheorie*, Bd. 2, *Die allgemeine Relativitätstheorie* (Braunschweig: Vieweg, 1921), p. 226.

predictions by Einstein's theory for where exactly the stars should appear in the sky during a solar eclipse; the results of the expeditions made Einstein into a global celebrity. For Oppenheimer and Snyder's work, however, a different kind of trajectory was important: in their case, essential was understanding the trajectories of particles or light rays that are flying straight at the center of one Schwarzschild's point-stars.

The consensus was that such a particle or light ray would travel ever closer to the star, but would never pass beyond the Schwarzschild radius; it should even take an infinite amount of time for the particle to reach that radius (see Figure 1). This would imply that, should a star implode, it would take forever before the star would have contracted to its Schwarzschild radius; the British astronomer Arthur Eddington spoke of a "magic circle",[3] on which light and matter aggregate, without ever passing this horizon.

Gradually, however, opinions changed. In 1932, the Belgian priest-cosmologist Georges Lemaître was doing calculations on relativistic models of the universe, and, more or less as an aside, he observed that the infinite value of the gravitational field at the Schwarzschild radius was not a mathematical necessity, but rather depended on the chosen perspective.[4] That is: according to the theory of relativity, one can make any choice for ruler or clock when describing physical phenomena—in other words: one can make any choice for the mathematical coordinates one uses to describe the location in space and time of a phenomenon. For certain observers, particular rulers and clocks will be a more natural choice, just as someone measuring up a soccer field will prefer using a land surveying instrument rather than a classroom ruler, or, e.g., someone on a railroad platform will prefer using a clock hanging there, rather than a clock whizzing by on a passing train. Formally, however, there should be no difference: the theory of relativity asserts that any set of coordinates, any choice of rulers and clocks is fully equivalent to any other for describing natural phenomena. Still, Lemaître's analysis entailed that when describing stars, researchers had so far implicitly always chosen the clock (or time coordinate) that was a natural pick for an observer resting at a great distance from the star. Yet, if one were to use the natural choice for the clock of an observer freely falling in the curved space of the star, it would become clear that that observer would not experience an infinitely strong gravitational field at the Schwarzschild radius.

---

[3] Arthur S. Eddington, *Space, Time and Gravitation*, Cambridge: Cambridge University Press, 1920, on p. 98.
[4] On Lemaître, see Jean Eisenstaedt, "Lemaître and the Schwarzschild solution", pp. 353-389 in John Earman, Michel Janssen and John Norton (eds.), *The attraction of gravitation. New studies in the history of general relativity*, Boston: Birkhäuser, 1993.

Lemaître wrote up his findings in the somewhat obscure local *Publications du Laboratoire d'Astronomie et de Géodésie de l'Université de Louvain* (Lemaître 1932), and they drew little attention. He also undertook extensive travels across North America, and here shared his ideas with his fellow cosmologists Howard Robertson and Richard Tolman, who were working in Princeton and at the California institute of Technology in Pasadena, respectively. Yet, for a number of years even for these specialists of the theory of relativity the full implications of Lemaître's results remained unclear. A visible publication that unequivocally stated that there was nothing unusual about the space near a Schwarzschild star in the perspective of an infalling observer did not materialize.

***

J. Robert Oppenheimer was born in 1904 in an affluent family on New York's Upper West Side. He grew up in an apartment filled with Picassos and was educated at the schools of the 'Ethical Culture Society': a secular Jewish group that emphasized personal growth through cultural development, and placed great emphasis on taking responsibility for society at large. After a formative trip through the deserts of New Mexico—Oppenheimer would forever remain enamored by its landscapes—he continued his studies at Harvard, supplemented by graduate study at the best European universities. In Leiden, he was nicknamed 'Opje', which means something like 'little one-up'; upon his return to the USA, this was anglicized to 'Oppie'.[5]

Robert Oppenheimer had a sensitive nature and suffered severe depressions. Anxious and paranoid, he once left a poisoned apple on his supervisor's desk during a troubled stay in Cambridge, England. The university was persuaded not to hand the matter over to the police, but only on the condition that Oppenheimer would seek psychiatric treatment. On the European continent, in Göttingen, Leiden and Zurich, he again found his intellectual self-confidence; so much so that, in fact, more than a few thought him arrogant.

It was clear that Oppenheimer was exceptionally talented, even though, as a relative newcomer, he only played a minor role in the quantum revolution of the 1920s. He settled down in California, where he combined positions at Caltech and Berkeley. He surrounded himself with an enthusiastic group of students in Berkeley and together they formed the first school of theoretical physics on the American continent. Oppenheimer often picked up his ideas in Pasadena, to then develop and distribute them among his numerous students in Berkeley. This also happened in the case of the black hole: Tolman shared his expert knowledge of the relativistic Schwarzschild spacetime with Oppenheimer, who, together with Snyder in Berkeley, integrated this with novel understanding of the physics of stellar evolution.[6]

---

[5] For biographical information on Oppenheimer see, Kai Bird and Martin J. Sherwin, *American Prometheus. The Triumph and Tragedy of J. Robert Oppenheimer*, New York: Vintage, 2006.
[6] On Oppenheimer's path to black holes, see e.g. Hufbauer 2005.

California's academic culture became increasingly leftist in the 1930s. Oppenheimer expanded his Berkeley group of friends with progressive intellectuals: the camaraderie in these circles made him feel connected with "the life of my time and my country."[7] He befriended Haakon Chevalier, a charismatic lecturer of French and in all likelihood a member of the American Communist Party. Physics, meanwhile, had turned its attention to the atomic nucleus. Hans Bethe, an immigrant from Germany, used the newly gained insights to explain the energy balance of stars: the fusion of hydrogen nuclei to helium nuclei, and from helium to carbon, and so on all the way up to heavy iron cores, fuels the radiative processes that prevent a star from collapsing under its own gravitational weight. Oppenheimer had found his subject: stellar evolution, studied using the general theory of relativity. He was ready to deliver on the promise of his prodigy.

***

Albert Einstein was born in 1879 and also published an article about the existence of black holes in 1939. Einstein's biography is familiar to many: after a fairly comfortable childhood in a middle class family of entrepreneurs and engineers, and an interrupted high school education in Munich, he enrolled at the novel *Eidgenössische Technische Hochschule* in Zurich, for a degree in physics. His first proper professional post was as patent officer in Bern, where he authored the five articles that together mark his 'miracle year', 1905. These papers seeded the revolutions of twentieth century physics: light sometimes behaves as a particle; and clocks may run slower while rulers may contract, given their relative state of motion compared to other clocks and rulers. Einstein subsequently began his ascent of the academic ladder until in 1914, he accepted a unique position in Berlin that awarded him substantial time and funds to just do research. In Berlin he finalized the 'general' theory of relativity: he expanded the scope of relativity to now also include the gravitational force. Einstein soon became a public figure, and not only because of his science: his efforts on behalf of democracy, Zionism, and particularly pacifist internationalism produced a near permanent presence in newspapers around the world.

Einstein no longer resided in Berlin in 1939. He had relocated to the USA following the National-Socialist seizure of power. It had been immediately clear to him at that moment that he could not return to Germany from a trip to America. He first stayed on the Belgian coast for close to a year to organize his relocation, guarded by Belgian security officers as the authorities feared for his safety. He had already been sounded out to join the Institute for Advanced Study in Princeton, NJ, and under these circumstances it was quickly decided that he would settle there.

One could argue that quantum mechanics came too late for Einstein: when the first formulation of the full quantum theory was attained in 1925, he had for a few years already been developing his own, alternative way to understand the physics of the very small. Einstein expected that by connecting up gravity and electromagnetism "in the mathematically most natural way"—by 'unifying' them—he would be able to offer a new foundation for all physical phenomena.[8] There is little doubt that he had been brought to this point of view by his own experiences and successes with the theory of relativity; in that case, mathematical innovation had been instrumental in attaining the result. He believed that a particle solution in such a unified, meta-generalized theory—a solution without infinities, unlike Schwarzschild's point-like star—might offer a good chance of describing the counterintuitive behavior of quantum particles; it might offer a fuller, more complete description of reality without having to resort to probabilities.[9]

---

[7] J.R. Oppenheimer to K.D. Nichols, 4 March 1954, available online at http://www.nuclearfiles.org/menu/library/correspondence/oppenheimer-robert/corr_oppenheimer_1954-03-04.htm (accessed on 27 December 2021).

[8] Albert Einstein to Walther Mayer, 23 February 1933, Einstein Archive Jerusalem, entry EA 18 163, cited on p. 242 in Jeroen van Dongen, 'Einstein's methodology, semivectors and the unification of electrons and protons', *Archive for History of Exact Sciences*, 58 (2004), 219-254. See also Albert Einstein, *On the Method of Theoretical Physics*, Oxford: Clarendon Press, 1933; John Norton, '"Nature is the realisation of the simplest conceivable mathematical ideas": Einstein and the canon of mathematical simplicity', *Studies in History and Philosophy of Science* B: *Studies in History and Philosophy of Modern Physics*, 31 (2000), 135-170.

Fellow physicists were initially awed by this project. Yet, they soon soured on it and by 1939, some even regarded Einstein's continued pursuit with slight embarrassment. His work appeared to have become entirely disconnected from the empirical, while it ignored the novel physics of cosmic particles and the atomic nucleus. To be sure, his reputation as a scholar of the gravitational force and the theory of relativity remained rock solid, yet some now regarded even this as having become a rather stale and dusty subject.

Robertson alerted Einstein to the novel interpretative challenges of Schwarzschild's gravitational field. Einstein quickly wrote up his own position on the matter, possibly alarmed that the infinities of Schwarzschild's solution may raise complications for his own program. He published a lengthy calculation in which he studied the gravitational field self-created by a sphere of particles that randomly rotated their center.[10] His conclusion: the particles would rotate at the speed of light if the radius of the sphere reaches too close to the Schwarzschild radius. Since particles can not move faster than light, the sphere can not be contracted beyond that point. So, the Schwarzschild radius "does not exist in physical reality […] because matter cannot be concentrated arbitrarily."[11] Einstein's particles remain forever rotating on their ideal trajectories—yet, had he not considered the possibility of falling straight down?

***

In the summer of 1939, Einstein was vacationing on Long Island. He was visited by three young Hungarian emigré scientists, some of whom he had known as students back in Berlin. It had just been announced that particular uranium isotopes fission if they capture a slow moving neutron, and that such isotopes release multiple neutrons in the process. One of the Hungarians, Leo Szilàrd, quickly realized that it would be possible to start a chain reaction with these isotopes: should one of the newly released neutrons again be captured by a uranium core, another fission would result. If certain parameters were tuned by enriching natural uranium, the whole process could produce an exponential chain reaction, and thus, an explosion: Szilàrd realized that one could construct a nuclear bomb. The richest uranium ores were found in the Congo, which was a Belgian colony; so, Szilàrd and his friends Edward Teller and Eugene Wigner had thought it wise to inform the Belgian government—and since Einstein was a good friend of the Belgian queen, they called on him on Long Island.

As they found themselves in the United States, however, they decided that the American government should really be informed first. Einstein dictated two versions of a letter to that effect and signed both. One was communicated to President Roosevelt, who was thus informed that it would be "possible to construct extremely powerful bombs of a new kind."[12] Roosevelt was further advised to get in touch with scientists who knew the physics of uranium chain reactions. Germany might already be gearing up to construct such a bomb, Einstein's letter suggested.

---

[9] For a discussion of Einstein's program, see Jeroen van Dongen, *Einstein's unification*, Cambridge: Cambridge University Press, 2010.
[10] Jeroen van Dongen and Dennis Lehmkuhl, 'Albert Einstein and the black hole', in preparation.
[11] Albert Einstein, 'On a stationary system with spherical symmetry consisting of many gravitating masses', *Annals of Mathematics* 40 (1939), pp. 922-936, on p. 936.
[12] Albert Einstein to F.D. Roosevelt, 2 August 1939, available online at https://www.atomicheritage.org/key-documents/einstein-szilard-letter (consulted on 23 December 2021).

Roosevelt was sufficiently impressed to quickly create the Advisory Committee on Uranium, whose first meeting was attended by Szilàrd, Teller and Wigner. Einstein chose to remain at a distance, and in any case would not easily have been invited to join the later effort to construct an atomic bomb, given his leftist positions and independent character (Jerome 2002). At first nothing particularly substantial took place towards the development of a weapon, until the British woke the Americans out of their slumber: Rudolph Peierls and Otto Frisch, Jewish refugees in Oxford, had calculated that only one kilogram of enriched uranium would be required for a working bomb. This was much less than earlier calculations had suggested and the 'Manhattan project' was soon started up.[13]

Oppenheimer was asked to serve as its scientific director. He was one of the few (perhaps only) American born physicists who was sufficiently informed about all the different aspects of the physics that the project would require. Furthermore, his reputation of genius, as well as his evident ambition and charm, made him the ideal candidate. Oppenheimer realized that any political activity may hamper his involvement with the project and he quickly decided he should be "cutting off every communist connection."[14] Oppenheimer was appointed director in October 1942, and a few months later he left for New Mexico. On a mesa in the middle of the desert a novel town, Los Alamos, was created. Here, in complete seclusion, work began on the creation of the new weapon.

Before Oppenheimer and his new wife Katherine Puening left Berkeley, they invited the Chevaliers over for a small dinner party. Haakon and Barbara Chevalier had become intimate friends, and they wanted to enjoy a quiet evening to say goodbye. Early in the evening, Oppenheimer withdrew to the kitchen to prepare drinks, and he was followed by Chevalier. Chevalier had recently met George Eltenton, a British physicist who was working for Shell oil in California. Eltenton had asked Chevalier to put a request to Oppenheimer. Would Oppenheimer be willing to share information about his new work with a Russian diplomat in San Francisco?

\*\*\*

J. Robert Oppenheimer published only three articles on stellar evolution. The papers had been prompted by Bethe's latest work, which he may have liked to have done himself; the first of the three articles, co-authored by Robert Serber, did not contain much more than a recommendation for Bethe's theory, although the paper also offered an estimate of the lowest value of the mass of a star's neutron core at which the star would still be stable. To do the calculation, Oppenheimer and Serber had still depended on the old gravitation theory of Newton.[15]

George Volkoff, one of Oppenheimer's students in Berkeley, had been thinking about stellar energy sources for at least a year by then. Now Oppenheimer wanted to know what happens when a heavy star runs out of resources to run nuclear fusion processes. To arrive at an accurate result, one really ought to use the general theory of relativity instead of Newton's theory, and together with

---

[13] On its history, see e.g. Rhodes (1986).
[14] Oppenheimer to A. Compton, cited on p. 184 in Bird and Sherwin.
[15] J. Robert Oppenheimer and Robert Serber, 'On the stability of stellat neutron cores', *Physical Review*, 54 (1938), 540.

Tolman, Oppenheimer and Volkoff drafted the equations that would need to be solved.[16] Thus, they returned to Schwarzschild's model.

Early in the 1930s, a Russian researcher, Lev Landau, had estimated that stars that are heavier than one and a half times the solar mass would implode to a neutron star, which consists only of closely packed neutrons. Oppenheimer and Volkoff showed that only neutron stars with a mass that was less than three quarters of a solar mass could be stable; the combined results implied that such stars would be quite rare. Most burnt out heavy stars would continue their contraction, passing through the neutron star phase without notice. What would then end up happening to these fully imploded stars?

Oppenheimer likely chose Hartland Snyder for his next project because of his mathematical acumen. The calculations were difficult, and once concluded, they produced "very odd" results, as Oppenheimer wrote Dutch physicist George Uhlenbeck.[17] In the perspective of an imaginary observer that remained at a large distance of a star, "the gravitational deflection of light […] will prevent the escape of radiation except through a cone about the outward normal of progressively shrinking aperture as the star contracts."[18] Only the gravitational field of the star remained, as a dark grave of what was once a ball of light. A message from behind the star's horizon could never reach across; inside the horizon, the sky would be alight, yet on the outside, the star remained dark and dead.

As we saw, earlier researchers believed that the contraction of the star would take forever. Oppenheimer, likely through interactions with Tolman and Robertson, realized that a second perspective was important: the perspective and time coordinate of an observer that falls in together with the contracting matter. According to that observer, Oppenheimer stated unequivocally, the required time before full "isolation" sets in, is "finite and may be quite short."[19] An essential difference with Einstein's account was that Oppenheimer and Snyder had focused on an imploding star, with matter collapsing radially inward, across the Schwarzschild radius. Oppenheimer was led to the question because of his interest in recent work on the nuclear physics of stars—a subject that Einstein was hardly aware of. In the end, the result was that Einstein's particles remained forever circling around the gravitational center, while Oppenheimer's fell straight down, into the infinite abyss of the black hole.

***

Oppenheimer immediately declined Chevalier's proposal. Piqued, he made clear that he did not wish to have any role in the suggested "treason".[20] He was not at all persuaded by the argument that a group of Washington reactionaries were withholding vital strategic information from Russian WWII allies that were fighting for their survival. Oppenheimer and Chevalier quickly returned to their dinner party and their wives, and Oppenheimer did not immediately report the exchange to his military superiors, even though, as incoming leader of the atomic bomb program, he was now living under strict security protocols. Perhaps he wished to protect his friend or did not consider the conversation to have been all that important.

---

[16] Richard Tolman, 'Static solutions of Einstein's field equations for spheres of fluid', *Physical Review*, 55 (1939), 364-373; J. Robert Oppenheimer and George Volkoff, 'On massive neutron cores', *Physical Review*, 55 (1939), 374-381.

[17] J. Robert Oppenheimer to George Uhlenbeck, 5 February 1939 on p. 209, in A.K. Smith and C. Weiner, *Robert Oppenheimer, Letters and Recollections*, Cambridge MA: Harvard University Press, 1980. In the same letter, in which Oppenheimer was also sharing his excitement on the recent uranium results with Uhlenbeck, he thought it "not too improbable that a ten cm cube of uranium deuteride […] might very well blow itself to hell."

[18] Oppenheimer and Snyder, p. 456.

[19] Oppenheimer and Snyder, p. 456.

[20] Oppenheimer, quoted in Bird and Sherwin, p.195.

Work on the mesa finally really got underway in March of 1943. Hans Bethe led the theory department, while Robert Bacher led the experimental group. Oppenheimer grew into his role and became a decisive leader who had a firm grip and good overview of the entire enterprise. His superior was general Leslie Groves, a straightforward and somewhat gruff authoritarian, who nevertheless developed a good working relation with Oppenheimer. One of Oppenheimer's best friends, Isidor Rabi, decided against joining the effort in Los Alamos: he held that "the culmination of three centuries of physics"[21] should not be a weapon of mass destruction. Oppenheimer however felt that one needed to develop the bomb in any case, before the Germans would.

There were two options to create a fission bomb: the first was to fire a uranium 'bullet' into another uranium mass; the second was to make a quantity of plutonium implode. In both cases, the exact right density of a particular isotope would result for a chain reaction to set in. The process to enrich the heavy metals so that they would have the right mix of isotopes was lengthy and expensive, and the scientists in New Mexico continuously feared a German success. There was also a third possibility: a hydrogen bomb. This would be yet more, and a lot more destructive than a fission bomb. In that case, the explosion would be the result of fusion processes, just as they took place in the interior of a star.

Edward Teller, one of the Hungarians that had visited Einstein, was very excited about the possibility to create such a 'superweapon'. Construction of a hydrogen bomb was however considered too great a challenge and was not assigned a high priority. When Teller began to neglect his responsibilities for the fission bomb, Oppenheimer reassigned those to Peierls; from that moment on, Teller could think as much as he would like on the 'super', but he would no longer hold up progress on the uranium and plutonium bombs. Teller felt snubbed, just as earlier, when Bethe and not he had been made leader of the theory group.

Germany had already been beaten when the fission bombs were ready to be tested, and it turned out that the German program had never been a serious competitor. Oppenheimer and the military leadership allocated part of the desert to test the bomb: they chose an area that Spanish colonizers had once called *Jornada del Muerto*, yet Oppenheimer now named it *Trinity*. Soon after, on 16 July 1945, the sky was filled with white light. "Now I am become death, the destroyer of worlds", Oppenheimer thought.[22]

***

Three weeks later, on 6 August 1945, Einstein joined his secretary Helen Dukas for tea. They were vacationing in the Adirondack mountains in upstate New York, and Einstein had just finished an afternoon nap. Dukas told him that the atomic bomb had been dropped on Hiroshima. "O my God", Einstein said.[23]

Einstein had not been involved with the American bomb program after he had sent his communication to Roosevelt; yet he would often be reminded that his letter had been its beginning. He argued that the importance of that letter should not be taken out of proportion—it still took quite some time and additional research for the program to get fully started up—nevertheless, Einstein also expressed that, despite his earlier pacifism, a vigorous self-defense against virulent National-Socialists had been absolutely necessary; and he further added the point that others had also raised about the danger of the Nazi's getting the bomb first. Still, Einstein did not think it had been right to use the weapon in Japan.[24]

---

[21] Rabi, according to Oppenheimer in his letter to Isidor I. Rabi, 26 February 1943, on p. 250 in Smith and Weiner.
[22] See p. 309 in Bird and Sherwin.
[23] Einstein, quoted in Walter Isaacson, *Einstein. His life and universe*, New York: Simon and Schuster, 2007.

He later wrote to a Japanese publication that resistance to particular weapons is pointless; one should rather focus on preventing war and reasons to go to war. Einstein was convinced that a 'world government' should be created, which would make individual armies superfluous. In his opinion, the United Nations fell short of what was necessary, as it was an assembly of nation states; these needed to be subordinated to a higher authority, which should be the only authority that held detailed knowledge of the workings of atomic weapons.

In 1946 Einstein became chairman of the *Emergency Committee of Atomic Scientists*. Its goal was to inform the American public of the dangers of nuclear weapons: there was no doubt that other countries would soon also have atomic bombs in their arsenal and no defense against such weapons was possible; they would certainly be used in a future war and destroy human civilization. The only solution was international control and, eventually, banishing all war. Einstein was convinced that the world would go to ruin without the installment of a world government.

Just before he passed away in April of 1955, Einstein addressed the world one last time in a manifesto that had been drafted together with philosopher Bertrand Russell. They warned of the dangers of the hydrogen bomb, which was a thousand times more powerful than the bomb that had been dropped on Hiroshima, and that had meanwhile been successfully tested. Einstein and Russell once again called for the banishment of all wars; as a first step, the fusion weapon should be banned. However, Einstein no longer had much influence nor easy access to the highest echelons of American political power. Some politicians and administration officials thought him subversive or some kind of crypto-Communist. Others, focused on securing the upper hand in the Cold War, saw Einstein increasingly as an unpragmatic idealist; someone who, at a safe Platonist distance from the center, preferred to remain running around in pointless circles.

***

Oppenheimer became director of the Institute for Advanced Study in 1947. This made him formally Einstein's boss. He had not found it easy to return in his old academic positions in California. In Princeton, on the East Coast, he would be close to Washington DC, where his advice was greatly valued and he enjoyed an enormous reputation as the 'father of the atomic bomb.' That same year he had been called upon to serve as chairman of the scientific General Advisory Committee, which assisted the Atomic Energy Commission (AEC). The AEC was the highest authority in the US government that decided on all nuclear research and its possible applications; it answered directly to the President and Congress.

Einstein and Oppenheimer were not particularly close. Einstein did not consider Oppenheimer's contributions to physics to be of the first rank, while Oppenheimer, who was completely convinced of the correctness of quantum mechanics, already judged Einstein in the 1930s as "completely cuckoo."[25] Nonetheless, Einstein valued Oppenheimer's qualities as an administrator, and Oppenheimer of course held Einstein's early work in the highest esteem. Yet, even though they were cordial with one another, Einstein could not understand Oppenheimer's positions: the latter feared an arms race as well, but chose not to voice his concerns too loudly or too publicly.

---

[24] On Einstein and the bomb, see e.g. Bernard T. Feld, 'Einstein and the politics of nuclear weapons', pp. 369-393 in G. Holton and Y. Elkana, *Albert Einstein. Historical and cultural perspectives*, Mineola NY: Dover, 1997.
[25] J. Robert Oppenheimer to Frank Oppenheimer, 11 January 1935 on p. 190 in Smith and Weiner.

Oppenheimer foremost wished to retain his influence. He did not always comport himself in the most equilibrious manner, however. In the fall of 1945, he was invited to an audience with President Truman, who was of the opinion that the Soviets would never succeed at building an atomic bomb. Oppenheimer tried to convince Truman to create international control mechanisms for nuclear technology—and, taken aback by Truman's lack of understanding of his position, he softly stated that he had blood on his hands. Truman was angered by the comment and let it be known he did not wish to see this "cry-baby scientist" again.[26]

Edward Teller had never given up his ambition to construct the 'super'. When it became clear in 1949 that the Soviets had developed a plutonium bomb, he saw an opportunity: the fusion weapon could now be sold as the best way to stay ahead of the USSR. Oppenheimer questioned the feasibility of a hydrogen weapon, and was distraught that many considered such a destructive weapon the best way to maintain peace. His Committee advised against the accelerated development of a fusion bomb: the weapon could not be deployed without committing genocide. In light of the technical and moral objections, it found that the security of the United States would not be served by the development of this weapon. Truman, however, felt that if the Russians could develop a hydrogen bomb, the Americans surely needed to do so as well.

In 1953 Oppenheimer lectured to a select audience of lawyers and bankers. He expressed his doubts about an American policy that resulted in ever bigger and more weapons to strike ever more forcefully against a Soviet Union that was arming itself at an equally alarming pace. He also criticized the secrecy in which all things nuclear had been shrouded. Lewis Strauss, prominent AEC member, was furious, and began undermining Oppenheimer's position. He made sure stories were published that made clear Oppenheimer had delayed the development of the hydrogen bomb, and had J. Edgar Hoover's FBI circulate a memo that questioned his loyalty. President Eisenhower immediately suspended Oppenheimer's security clearance: a "blank wall" was to come down around him.[27]

Strauss presented Oppenheimer with two options: either he would quietly step down from his position as government councilor, or face a formal hearing, in which it would be decided whether his security clearance could be reinstated. Oppenheimer chose to fight the allegations. Einstein advised him not to face the witch-hunt but to resign—when Oppenheimer told him he would not, Einstein called him a "*Narr* [fool]" to an assistant.[28] During the interviews, which took many long days, Edward Teller was asked to testify about Oppenheimer's resistance to the hydrogen bomb. Teller did not believe that Oppenheimer would be disloyal to the United States. Yet, he often found it difficult to understand his actions, considered them confused and complicated, and he saw "the vital interests of this country" rather in the hands of someone he could more readily trust.[29]

Oppenheimer's kitchen conversation with Haakon Chevalier also came up during the hearing. He had eventually himself informed the military about the conversation in August of 1943, and later the FBI. Initially, he had omitted mentioning Chevalier's name, but offered it up later. His interrogators held a transcript of his first interview on the matter, and it now turned out that Oppenheimer had added all kinds of made up details, without there having been any obvious reason to do so: multiple scientists would have been approached, mention would have been made of microfilm as a way to transfer the information. Oppenheimer could not explain why he had added all these untruths. He could only admit that he had lied; he even admitted that he had "told a whole fabrication and tissue of lies."[30] Robert Oppenheimer fell; he fell straight down.

---

[26] Truman, in Bird and Sherwin on p. 332.
[27] Eisenhower, in Bird and Sherwin, on p. 480.
[28] Einstein to Bruria Kaufmann, in Bird and Sherwin, on p. 495.
[29] Edward Teller, cited on p. 710 of The United States Atomic Energy Commission, *In the Matter of J. Robert Oppenheimer, Transcript of Hearing before Personnel Security Board and Texts of Principal Documents and Letters*. Cambridge MA: The MIT Press, 1971.

---

[30] Roger Robb (when interrogating Oppenheimer, who answered affirmatively), quoted on p. 149 of *In the Matter of J. Robert Oppenheimer*.